# Preserving the beamforming effect for spatial cue-based pseudo-binaural dereverberation of a single source

Sania Gul[1], Muhammad Salman Khan[2], Syed Waqar Shah[1]

[1]Department of Electrical Engineering, University of Engineering and Technology Peshawar, Pakistan.

[2]Department of Electrical Engineering, College of Engineering, Qatar University, Doha, Qatar.

**Abstract_** *Reverberations are unavoidable in enclosures, resulting in reduced intelligibility for hearing impaired and non-native listeners and even for the normal hearing listeners in noisy circumstances. It also degrades the performance of machine listening applications. In this paper, we propose a novel approach of binaural dereverberation of a single speech source, using the differences in the interaural cues of the direct path signal and the reverberations. Two beamformers, spaced at an interaural distance, are used to extract the reverberations from the reverberant speech. The interaural cues generated by these reverberations and those generated by the direct path signal act as a two-class dataset, used for the training of U-Net (a deep convolutional neural network). After its training, the beamformers are removed and the trained U-Net along with the maximum likelihood estimation (MLE) algorithm is used to discriminate between the direct path cues from the reverberation cues, when the system is exposed to the interaural spectrogram of the reverberant speech signal. Our proposed model has outperformed the classical signal processing dereverberation model 'weighted prediction error' in terms of cepstral distance (CEP), frequency weighted segmental signal to noise ratio (fwsegSNR) and signal-to-reverberation modulation energy ratio (SRMR) by 1.4 points, 8 dB and 0.6 dB. It has achieved better performance than the deep learning based dereverberation model by gaining 1.3 points improvement in CEP with comparable fwsegSNR, using training dataset which is almost 8 times smaller than required for that model. The proposed model also sustained its performance under relatively similar unseen acoustic conditions and at positions in the vicinity of its training position.*

## 1. Introduction

The human auditory system has a spectacular capability to analyze, process and select a particular signal from a complex acoustic environments, significantly so from the signals contaminated by room reflections. This ability is attributed to a large extend to still highly enigmatic and complex auditory and cognitive mechanisms, which rely on the binaural signals, which enable the listeners to analyze the acoustic scene and suppress the undesired signal components [1].

----------------

*Corresponding author: Department of Electrical Engineering, College of Engineering, Qatar University, Doha, Qatar. *Email address:* salman@qu.edu.qa

When sound is emitted from a source, it reaches the listener by two means. The first is by the direct path between the source and the listener, the other is by the reflections from the surrounding objects and walls. These reflections are called reverberations. Reverberations are delayed and decayed replicas of the direct path sound. They are delayed, as their propagation path is longer than the direct path. They are decayed, as they are absorbed by the objects from which they are reflected and by the medium through which they travel.

Reflections that arrive shortly after the direct wave are called early reflections (within 10 to 50ms of direct wave) and those that arrive after this time are called late reflections.

In an anechoic environment, where there is only the direct sound, a normal hearing listener can accurately localize arbitrary sound sources due to the presence of unambiguous spatial (interaural) cues (the interaural time difference (ITD) and the interaural level difference (ILD) cues), and the spectral cues that are provided by the interaction of pinnae, head and torso with the sound field [1].

In case of reverberant environment, the direct sound is accompanied by reverberation. As reverberations can be viewed as the same source signal coming from several different sources (virtual) placed at different locations in an enclosure [2], the spatial cues of reverberations are different from those of the direct waves [3], simulating the effect of many sources, as shown in Figure 1 below. While the early reflections have a specular nature, the late reverberations generally do not show specular behavior and are somewhat diffuse in nature [4].

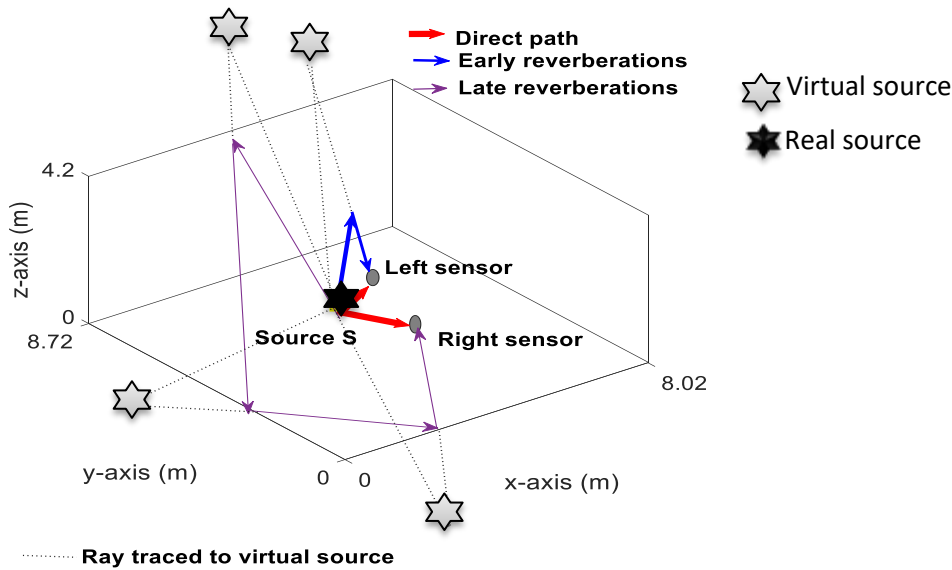

**Figure 1:** Reverberations and the creation of virtual sources by them.

Reverberations create spaciousness to sound [2]. The sense of 'space' created by reverberations adds greatly to the realism and often makes the recorded music more enjoyable and attractive. Early reverberations add to the intelligibility of speech. If reverberations are present everywhere and are sometimes helpful, then why we do want to remove reverberations? The answer is dependent on the application. Reverberations result in distortion of the auditory cues and, typically, lead to the reduced performance, for instance, in localization [1].

In the case of speech, late reverberations are detrimental to the intelligibility. Although displeasing for normal hearing people, its effect on speech intelligibility is especially noticeable for non-native listeners and for hearing impaired persons using assistive listening devices (e.g. cochlear implants (CI) and hearing aids) [5]. Other perceptual effects of reverberations are the 'box effect' and the 'distant talker effect' [2].

Beamforming has long been used to combat reverberations and unwanted noise generated by other sources. Localization of sound sources in a number of audio conferencing applications is achieved by beamforming [3]. In beamformer, the signals obtained from the array of microphones are combined in such a way that the speech coming from the desired direction is enhanced, and noise or interference coming from other directions is attenuated [6] and [7]. The spread of the beam is controlled by the operating frequency, geometry, inter-microphone distance and the number of microphones in the beamformer array. However, the beamformers can only partially suppress reverberation because reflections coming from the look direction are not attenuated [8].

Recent years have witnessed an increase in hands-free communication. The increase in the use of portable devices accompanied by the expansion of broadband internet access paved the way for many new applications e.g. teleconferencing, automatic speech to text conversion, voice controlled device operation, speaker identification, source separation, car interior communication system and sophisticated assistive listening devices, all of which demand high quality speech without being contaminated by reverberations and noise [2], as machine listening is not as resilient as human listening.

As human and animals have two ears, so speech enhancement and source separation in applications such as assistive listening devices and binaural robot audition requires direction-of-arrival (DOA) estimation of a sound source by utilizing the binaural cues. The ever increasing use of headphones or earpieces e.g. in binaural telephony, teleconferencing, hands-free devices and interfaces, immersive-audio rendering, and so on demands binaural dereverberation as an essential preprocessing step in order to ensure reception comparable or better to that of normal listening. However, binaural dereverberation is not a trivial task. Apart from the challenging task of reducing reverberation without introducing audible artifacts, binaural dereverberation should preserve the interaural cues, because it has been shown that bilateral signal processing can otherwise adversely affect source localization [1].

Although the classical single channel dereverberation algorithms using signal processing (e.g. spectral subtraction, prediction error and inverse filtering) can be extended for the binaural use [1], special binaural dereverberation techniques are also proposed, where the reverberant speech signal is decomposed in time frequency (TF) domain in order to suppresses components that are estimated to be mainly reverberant [9]. The binaural dereverberation models of [9] and [10] use the interaural coherence (IC) between the binaural channels at every TF unit to estimate the post filter gain. In binaural source localization model of [11], the dereverberation step is achieved by carrying out the direct-path dominance (DPD) test on each TF bin in order to ensure that only the TF units dominated by the direct path are used for direction of arrival (DOA) estimation of source. In [12], spectral subtraction is proposed to suppress late-reverberation for the binaural signals [1], later adopted by [13] and [14]. Filtering is also employed for binaural dereverberation [15].

With the development of Neural Networks (NNs), there has been tremendous improvement in a variety of speech recognition and acoustic signal processing tasks [16]. The binaural dereverberation models in [8], [16] and [17] uses artificial neural network (ANN) for binaural dereverberation preprocessing, the model in [18] uses the recurrent neural network (RNN) and interaural cues for speech enhancement in reverberant noisy conditions, while the models in [19] and [20] use the U-Net (a deep convolutional neural network (CNN)) for dereverberation, but these are monaural models.

In this paper, we use the human approach of dereverberation based on binaural cues' differentiation to separate the direct path signal from the reverberations. For this purpose, beamforming at the front-end, supported by the deep learning at the back-end is utilized. Beamforming is used for separating the echoes and the direct path signal from the reverberant speech. The direct path signal's spatial spectrograms and the extracted echoes' spatial spectrograms from beamforming are then used for the training of U-Net; a deep convolutional neural network. Due to the enormous learning capabilities of deep neural networks (DNNs), once trained, U-Net accompanied by machine learning, can separate the direct path speech from the reverberations, when presented with the binaural reverberant speech of a single source. In order to make sure that the trained system can work with binaural spatial cues in the field, the sensors are arranged in such a way that the system is trained on the binaural cues of both classes. To the best of our knowledge this is the first time U-Net is being used for the task of binaural dereverberation.

The rest of the paper is organized as follows. In the next section, we give an overview of the work implementing binaural dereverberation. We give our proposed system's overview in Section 3. In Section 4, we describe the experimental setup, the evaluation criteria and the comparison methods. We present experimental results and comparison statistics of different models in Section 5. We conclude the paper in Section 6.

## 2. Related Work

Although U-Net is a deep neural network (DNN), mostly used for image segmentation, we investigated U-Net for source separation in [21] and [22] and found the efficacy of this network for separating the sources on the basis of the differences in their interaural cues. The binaural audio mixture was converted from one dimensional (1D) time domain audio signal to two-dimensional (2D) interaural spectrogram in the TF domain. This 2D representation can be treated as an ordinary gray scale image by U-Net.

However, in this paper instead of separating different real sources from an audio mixture, we will separate the TF units of the real source and the virtual sources from a reverberant speech. The separation of direct path from reverberation is carried out on the basis of the differences in their interaural cues (as depicted in Figure 1). As mentioned in [21], unlike the ordinary gray image pixel intensity, the interaural parameters are not limited to the fixed range of 0 to 255. Also, there is no correlation between the neighboring TF units of mixture's interaural spectrogram that is known to exist among the neighboring pixels of an image.

In [21], U-Net was trained on interaural cues of two sources, which act as two separate classes, each having its own ground truth (GT). The GT in [21] is an image of all white or black pixels (one for either class), with dimension equal to the input interaural spectrogram. However, in this paper instead of the two speech sources, the interaural spectrograms of direct path (DP) and reverberations (REV) will act as two separate classes with the GT's design similar to the one used in [21].

As the U-Net failed to cluster the interaural phase difference (IPD) cues successfully in [21], and was replaced by the expectation maximization (EM) algorithm (a machine learning algorithm) in [22], so here we will train only one U-Net on the ILD spectrograms and use the maximum likelihood estimation (MLE) algorithm; a machine learning algorithm using initialization parameters of each class (DP and REV) as proposed in [23] for clustering the IPD cues. We have trained the U-Net from scratch and did not utilize any of the pretrained models e.g. ResNet or VGG, as none of them supports the input image size, we require for our proposed model. Resampling of the audio signal or its spectrogram, and varying the STFT window size could potentially be explored to adjust the input image size required for the pretrained networks, this will be investigated in future and is out of scope of the current study.

The interaural based source separation model of [23] utilizes the expectation maximization (EM) algorithm to cluster the ILD and IPD cues of each source. The EM algorithm is a maximum likelihood estimation (MLE) iterative algorithm, sensitive to initialization. It uses PHAT algorithm [24] for IPD initialization. An interesting feature of this model is the creation of an imaginary source called the 'garbage source (GS)' for dealing the reverberations. While the direct-path signal has interaural cues consistent with the specific direction of the source, reverberations have a diffuse character that may not fit a Gaussian source model particularly well [23]. The GS is assumed to have a uniform distribution, while the distribution of the real sources is assumed to be Gaussian. The interaural cues which are not nearer to the mean value of the real sources (those generated mostly by the reverberations) are assigned to the GS, resulting in performance improvement of the real sources, which would be otherwise contaminated, if these outliers are forcefully included in them. We will also use this concept in our proposed methodology

After having briefly introduced the existing binaural dereverberation models and the techniques from the existing source separation models, utilized in our proposed dereverberation model, we will discuss our methodology in the next section.

### 3. Proposed Methodology

We have named our proposed model 'Beamforming-Echo-Net (BENET)' as it tackles the echoes using the beamforming at front-end and U-Net at back- end. The block diagram of our proposed model is shown in Figure 2.

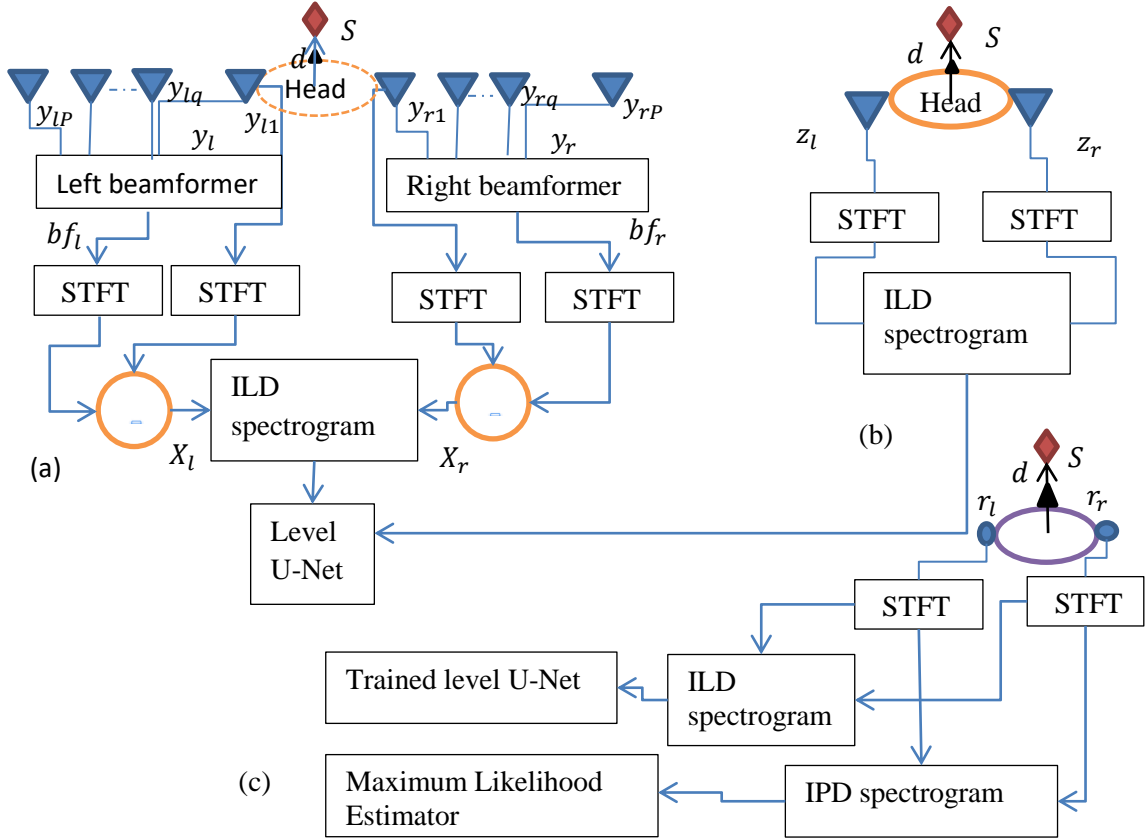

**Figure 2:** Block diagram of BENET. (a) and (b) shows the training phase, (c) shows the prediction phase. The dotted head in (a) shows that the first microphones of the left and right beamformers are separated by an interaural distance. The bold lined heads in (b) and (c) shows the binaural microphone setup.

The working of the model is explained as follows. Assume a single source $S$ is placed in an enclosure at a distance $d$ from the center of the two beamformers as shown in Figure 2 (a). The two beamformers are separated from each other by the distance equal to the distance between the two human ears. Except for this source, no other active source is present in the room. Assuming $P$ number of microphones in each beamformer array, the time domain signal $y_k$ at the input of each beamformer from all of its microphones is given as

$$y_k(n) = \{y_{k1}(n), y_{k2}(n), \dots, y_{kq}(n) \dots \dots \dots y_{kP}(n)\} \quad (1)$$

Where $k = \{l, r\}$ shows the left and right beamformers, n is the discrete time index, and $y_{k1}$, $y_{k2}$, ….$y_{kP}$ are the reverberant noisy speech signals collected respectively at the first, second….. and the $P^{th}$ microphone of the $k^{th}$ beamformer array. The signal $y_{kq}(n)$ is given as

$$y_{kq}(n) = h_{kqs}(n) * S(n) \quad (2)$$

where $h_{kqs}$ is the room impulse response (RIR) between the source $S$ and the $q^{th}$ microphone of the $k^{th}$ beamformer array, and * represents the convolution operation. The reverberant signal $y_k$ consists of direct path signal and the reverberations. The reverberations are separated from the direct path signal by taking the difference between the short time Fourier transforms (STFTs) of the reverberated speech $y_{k1}(n)$ collected at the first microphone and the output $bf_k$ of the $k^{th}$ beamformer, as shown in equation (3)

$$X_k(t,f) = Y_{k1}(t,f) - BF_k \qquad (3)$$

where $Y_{k1}$ is the STFT of $y_{k1}(n)$, $BF_k$ is the STFT of $bf_k(n)$ and $X_k$ is the STFT of the non-target (reverberations) output $x_k(n)$ of $k^{th}$ beamformer and $t$ and $f$ are the time and frequency indices in the STFT domain. The STFT conversion for equation (3) is performed by parameters listed in Section 4(h). The greater the number of microphones in the beamformer array, the more perfect would be the separation of the target (DP) and non-target (REV) sources. These time frequency (TF) domain reverberation signals from the left and the right beamformers are converted back to time domain signal and stored as separate left and right side reverberation audio files. These audio files will be used to create interaural spectrograms of reverberations (REV), which will behave as samples of the first class of the dataset used for the training of level U-Net (used for clustering ILD cues).

The second class of the dataset used for the training of level U-Net consists of the ILD spectrograms of the direct path signal (DP). They are obtained by assuming the source $S$ placed in anechoic conditions as shown in Figure 2(b). The source $S$ is considered to be at the same position as in Figure 2(a) but now in front of a binaural microphone setup. The direct path signal $z_k$ at the input of each microphone is given by

$$z_k(n) = h_{kS}(n) * S(n) \qquad (4)$$

where $h_{kS}$ is the room impulse response (RIR) between the source $S$ and the $k^{th}$ microphone of the binaural setup, and * represents the convolution operation.

The time domain direct path signal $z_k$ is converted to time frequency (TF) domain signal $Z_k(t,f)$ using the STFT parameters of Section 4(g).

$$Z_k(t,f) = \mathcal{F}(w(n) z_k(n)) \qquad (5)$$

where $\mathcal{F}$ is the STFT operator and $w(n)$ is the hamming window function given as $w(n) = 0.54 - 0.46 \cos(2\pi n/N), 0 \leq n \leq N, where\ N = L - 1$, and $L$ is the window length.

Then the interaural spectrograms are created by using equation (6) and (7).

$$\frac{Z_l(t,f)}{Z_r(t,f)} = \alpha(t,f) e^{i\phi(t,f)} \qquad (6)$$

where $Z_l(t,f)$ and $Z_r(t,f)$ are the STFT of signals at the left and right microphones respectively, $\alpha(t,f)$ is the ILD and $\phi(f,t)$ is the IPD at a time frequency (TF) point of the interaural spectrogram having discrete frequency index $f$ and time frame index $t$.

Equation (7) is used to convert the ILD $\alpha(t,f)$ to decibels (dB) at each TF point.

$$ILD(dB) = 20log_{10}\alpha(t,f) \qquad (7)$$

Using the same procedure of direct wave conversion, the audio files of left and right side reverberation (non-target signals) obtained from equation (3) are also converted to the interaural spectrogram using the STFT parameters of Section 4(g).

The ILD spectrograms of the DP and REV are used to train the level U-Net. Each class is accompanied by its respective GT mask. We have used MLE algorithm for clustering the IPD cues of the two classes. However, as MLE is a machine learning algorithm, it does not require any training before operation.

After training the level U-Net, the beamformers are removed. During the prediction phase, the network is exposed to the spectrogram of reverberant noisy speech $r_k(n)$, collected at each microphone of the binaural setup (Figure 2(c)).

The signal $r_k(n)$ is a composite signal, consisting of direct path signal $z_k(n)$, reverberations $x_k(n)$ and the random noise $rn$, each having its own distribution. The signal $r_k(n)$ is represented as in equation (8)

$$r_k(n) = z_k(n) + x_k(n) + rn \qquad (8)$$

The left and right channels are converted from time domain to TF domain by taking their STFTs and then the ILD and IPD spectrograms are obtained by applying equations (5), (6) and (7) on $r_k(n)$ using the STFT parameters of Section 4(h). The ILD spectrogram is given as an input to the trained level U-Net, which gives the probabilistic masks for each class at its softmax layer.

The IPD spectrogram is given as an input to the MLE block. The MLE algorithm requires three main parameters before it can start working. These are a) number of sources present in the speech, b) distribution of each source and c) the initial mean value of each source.

Under noiseless conditions, we have two main sources in speech; i.e. the direct wave and the reverberations. The distribution of IPD cues of the direct wave $z_k$ is assumed to be Gaussian as suggested in [23] and as reverberations $x_k$, are not clustered around any mean value, they are modeled as GS (again inspired by [23]) with uniform distribution. BENET also uses PHAT [24] algorithm for the initialization of mean value of the Gaussian distribution of the direct wave and a standard deviation of ±1 sample around that mean value. The uniform distribution for the reverberations is initialized to have zero mean value across frequency and a standard deviation of ±9 samples as in [23].

The observed IPD values from the reverberant speech i.e. $\angle \frac{R_l(t,f)}{R_r(t,f)}$ at each TF point (obtained from equation (6)) do not always map to the correct interaural time difference (ITD) due to spatial aliasing. So a top down approach inspired by [23] is used for the calculation of IPD, where IPD is estimated by plugging in different values of ITD ($\tau$) in the range from -15 to 15 samples in 0.5 sample increments. At the sampling frequency of 16 kHz, this range corresponds to around ± 1 milliseconds in 31.25 microseconds increments. The $\tau$ which produces the closest match to the observed IPD is selected. However, it is required that the delay ($\tau$) and the length of RIR must be smaller than the STFT frame length. Any portion of RIR above the STFT frame length would be treated as noise.

The phase residual error $\hat{\phi}$ is defined as the difference between the observed IPD and the estimated IPD and given in equation (9) as:

$$\hat{\phi} = \angle \frac{R_l(t,f)e^{-j2\pi f\tau}}{R_r(t,f)} \qquad (9)$$

$\hat{\phi}$ lies in the interval {-π, π}. The IPD residual for the direct wave is modeled as normal distribution. Let $\xi(\omega)$ and $\sigma^2(\omega)$ be the mean and variance of the IPD residual ($\hat{\phi}$). Then the IPD model for the direct wave at each TF point is given in (10) as:

$$p(\hat{\phi}(t,f;\tau)|\hat{\theta}) = \mathbb{N}\left(\hat{\phi}(t,f;\tau)\big|\xi_{z,\tau}(f),\sigma_{z,\tau}^2(f)\right) \qquad (10)$$

$\mathbb{N}$ represents Gaussian distribution. The subscripts with mean and variance symbols in equation (10) show that the IPD parameters of direct wave are dependent on both frequency $f$ and delay $\tau$. And

$$\hat{\theta} = \{\xi_{z,\tau}(f), \sigma_{z,\tau}^2(f), \psi_{z,\tau}\} \qquad (11)$$

represents all model parameters for the direct wave. $\psi_{z,\tau}$ is the mixing weight; i.e. the proportion of the total TF points of mixture belonging to direct wave at delay $\tau$. Using estimates of $\tau$ for source $S$ from PHAT-histogram [24], $\psi_{z,\tau}$ is initialized with the mean at each cross-correlation peak and standard deviation of ±1 samples [23]. The log likelihood $L$, given the observation $\hat{\theta}$, over all TF points is given in (12) as:

$$L(\hat{\theta}) = \sum_{t,f} \log p(\phi(t,f)|\hat{\theta}) \qquad (12)$$

For reverberation (REV) class, the IPD model at each TF point is given as

$$p(\hat{\phi}(t,f;\tau)|\hat{\Theta}) = U\left(\hat{\phi}(t,f;\tau)\big|\xi_{x,\tau}(f),\sigma_{x,\tau}^2(f)\right) \qquad (13)$$

where $U$ represents uniform distribution and

$$\hat{\Theta} = \{\xi_{x,\tau}(f), \sigma_{x,\tau}^2(f), \chi_{x,\tau}\} \qquad (14)$$

represents all model parameters for REV. $\chi_{x,\tau}$ is initialized as

$$\chi_{x,\tau} = 1 - \psi_{z,\tau} \qquad (15)$$

Marginalizing over all sources and all delays, the log likelihood function for DP is given as in (16).

$$L(\hat{\theta}) = \sum_{t,f} \log \sum_{x+z,\tau} [\mathbb{N}\left(\hat{\phi}(t,f;\tau)\big|\xi_{z,\tau}(f),\sigma_{z,\tau}^2(f)\right) \cdot \psi_{z,\tau}] \qquad (16)$$

The maximum likelihood solution is given as in (17).

$$L(\hat{\theta}) = max_\theta \sum_{t,f} \log p(\phi(t,f)|\hat{\theta}) \qquad (17)$$

The likelihood $v$ of each TF point belonging to direct path $z$ and delay $\tau$ is given as in (18)

$$v_{z,\tau}(f) \propto \psi_{z,\tau} \cdot \mathcal{N}\left(\hat{\phi}(t,f;\tau) \middle| \xi_{z,\tau}(f), \sigma_{z,\tau}^2(f)\right) \quad (18)$$

While for the REV, the probability of each TF point belonging to source $x$ is given as in equation (19).

$$\mu_{x,\tau} = 1 - v_{z,\tau}(f) \quad (19)$$

At the end, the probabilistic IPD mask for the direct wave is given as in equation (20)

$$M_z(t,f) = \sum_\tau v_{z,\tau} \quad (20)$$

And the probabilistic IPD mask for the reverberation is given as in (21)

$$M_x(t,f) = \sum_\tau \mu_{x,\tau} \quad (21)$$

As we do not want reverberations to be included in direct path signal at any cost, so as a precautionary measure, the initial standard deviation of DP signal is kept much smaller than that of REV (±1 vs. ±9).

The ILD soft mask from level U-Net and the IPD soft mask from MLE algorithm for the DP are combined to form product mask as given by equation (22).

$$Product\ mask = [Product\ of\ ILD\ and\ IPD\ masks\ of\ direct\ path\ (0\ to\ 8\ kHz)] \quad (22)$$

This product mask is applied separately on the left and right mixture's spectrograms $R_l$ and $R_r$, and the results are added together and converted back to time domain signal by taking the inverse short-time Fourier transform (ISTFT) in order to retrieve the target, which is then evaluated against the direct path signal collected at the first microphone of the left beamformer.

*Algorithm summary*

```
1) Task:   Single source dereverberation
2) Input: Reverberated speech source
3) Output: Dereverberated speech source
4) Convert the reverberated speech $r_k$ into interaural
   spectrogram using equations (5), (6) and (7).
5) Load pre-trained level U-Net.
6) Input the ILD spectrogram to level U-Net.
7) Save the outputs of softmax layer of the level U-Net.
8) Input the IPD spectrogram to MLE block (equations (8) to
   (21)).
9) Prepare the product mask (equation (22)).
10)      Apply the product mask to the STFT of the reverberated
   speech to retrieve the dereverberated speech.
```

## 4. Experimental Evaluation Parameters

The experimental setup, including the room layout, dataset, room impulse responses (RIRs), evaluation metrics, model parameters, and the overview of different dereverberation algorithms used for the comparison of our model is given below.

a) **Room Layout**

We will perform all our experiments inside simulated rooms with the source and the microphone setup during the training and the prediction phases, as shown in figures 3 (a) and 3 (b) respectively.

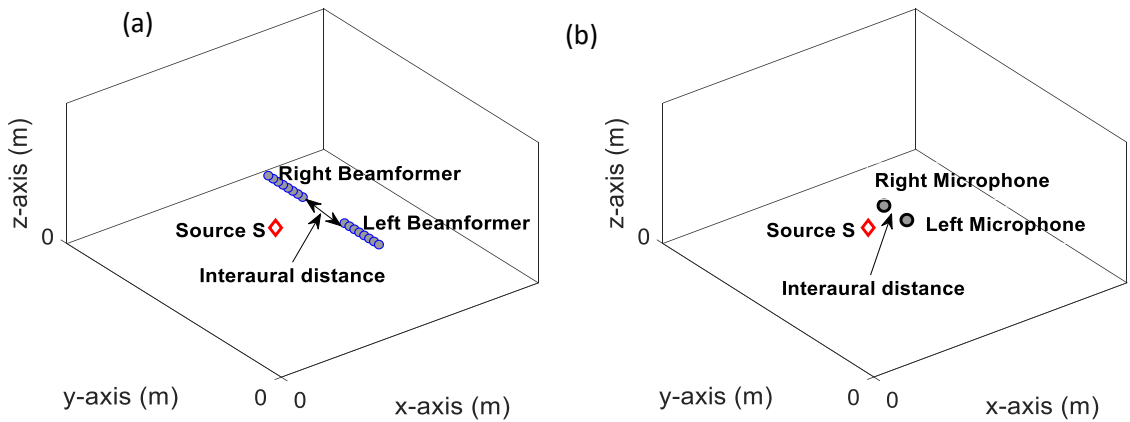

**Figure 3:** Room layout. (a). During training phase, (b). During prediction phase.

Each beamformer consists of 8 microphones arranged in a linear geometry with inter microphone spacing of 5 cm. In order to make sure that the trained system can work with binaural spatial cues in the field, the left and right beamformers in Figure 3(a) and the left and right sensors in Figure 3(b) are spaced by a distance equal to the interaural distance. This distance is equal to the average human interaural distance i.e. 0.17m. The size of room and the positions of the source and the microphones vary for different acoustic conditions (different $RT_{60}$ values) as shown in Figure 4.

### b) Room impulse responses (RIRs)

As the required RIRs for real rooms are not available in the online repository of RIRs, nor do we have the arrangements to record them ourselves, we have utilized the synthetic room impulse response of [25] for the training and prediction phase of BENET. Although this RIR generation toolkit is meant for moving sources, we have used it for a stationary source by setting the same starting and ending points for the source trajectory.

The RIR for binaural dereverberation should ideally include the head related impulse response (HRIR) added to $h_{ks}$. HRIR is a response that characterizes how an ear receives a sound from a point in space. As sound strikes the listener, the size and shape of the head, ears, ear canal, size and shape of nasal and oral cavities, density of the head, all transform the sound, boosting some of its frequencies and attenuating others and affecting the way it is perceived by the listener. However, as HRIR is required to be recorded in an anechoic room on the bare ear of a mannequin, but as neither the anechoic chamber nor the mannequin was available to us, so HRIRs are not included in the RIRs generated by simulation for the experiments of this paper. The RIRs used here are binaural in the sense, that the separation between the two beamformers and the two microphones (in the training and testing phase respectively) is kept equal to the interaural distance. But HRIR are not included in these RIRs. So, the system can be called as 'pseudo-binaural', and can be converted to truly binaural, just by adding HRIR in the RIRs.

The parameter settings for different rooms are given in Figure 4.

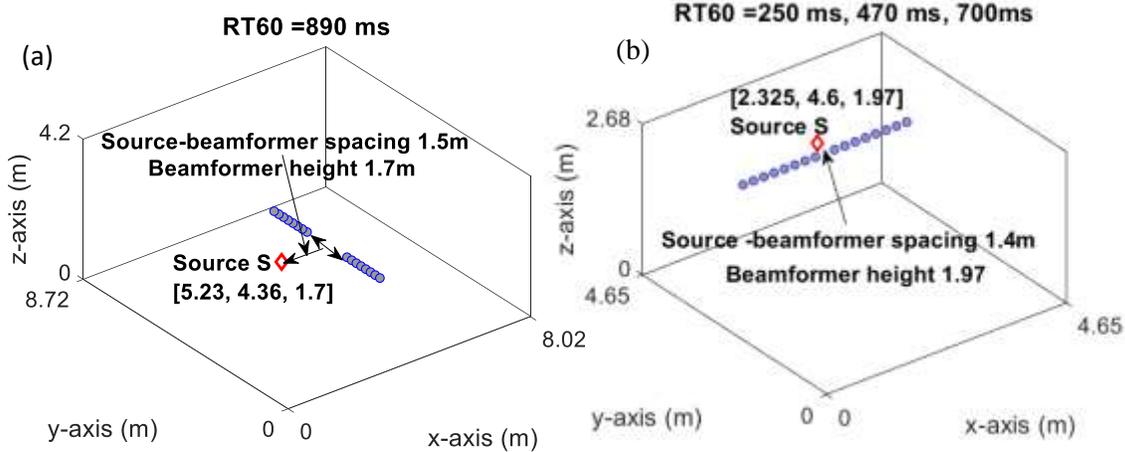

**Figure 4:** (a) and (b). Different RIRs and their respective room layouts during the training phase.

In both rooms of Figure 4, the absorption weight coefficients 'abs_weights' setting is kept as [0.6 0.9 0.5 0.6 1 0.8] (the default setting in ISM_setup function of [25]). These coefficients simulates a carpeted floor, and sound-absorbing material on the ceiling and the second x-dimension wall. For recording the direct path sound, the absorption weight coefficients 'abs_weights' setting is kept as [1 1 1 1 1 1], which leads to uniform absorption coefficients for all room boundaries. The direct path signal to the left microphone, inside an anechoic room, is used as reference for the performance evaluation. Expect for $RT_{60}$ =890ms, the room dimensions and the equipment setup for $RT_{60}$ = 250, 470 and 700ms reverberation times is similar.

### c) Dataset

The dataset used for audio sources is TIMIT database [26] and University of McGill database (TSP Lab) [27]. The training data specifications for level U-Net are given in table 1.

**Table 1:** Training data specifications

| | |
|---|---|
| Number of speakers from TIMIT and McGill database | 120 |
| Total audio files generated from each speaker to be used for level U-Net training | 100 |
| Audio file format | .wav |

Due to the restriction of spectrogram size to be equal to $2^D$ (where 'D' is the encoder depth of U-Net) at the input layer of U-Net, the duration of the clean source files used to create the ILD spectrograms (data samples for training the level U-Net) from the beamformer is different for different acoustic conditions, as listed in table 2.

**Table 2:** Audio file sizes used for creating the training data spectrograms in different acoustic conditions

| $RT_{60}$ (ms) | Clean file duration from [26] or [27] in sec | Reverberant file duration (after convolution with RIR) in sec | Reverberant file duration (after convolution with RIR) at sampling frequency of 16 kHz (in samples) | STFT spectrogram size (length (L) × width (W)) |
|---|---|---|---|---|
| 890 | 1 | 1.645 | 26320 | 1024 × 100 |
| 700 | 0.98 | 1.505 | 24080 | 1024 × 92 |
| 470 | 1 | 1.345 | 21520 | 1024 × 82 |
| 250 | 1 | 1.165 | 18640 | 1024 × 70 |

As the size of images of both classes given to any U-Net must be same, so the duration of direct path files created by the clean audio files of [26] and [27], after convolving them with the RIRs of $RT_{60}$ = 0ms in these rooms, is kept equal to that shown in column 3 of table 2.

However, after training, the prediction phase of U-Net is not subjected to such restrictions [21]. So during the prediction phase, five clean files of [26] (not used during the training phase), each of 2 seconds duration are convolved with the binaural RIRs of each room, to create the testing samples of reverberant speech. The process of convolution increases the duration of the clean speech beyond 2 seconds, but in order to keep the reverberations intact, we would not clip it back to 2 seconds. It will not pose any problem to the trained level U-net or the MLE algorithm.

### d) U-Net architecture

We have used Matlab (2019) U-Net architecture for our proposed system. The network architecture is shown in Figure 5. No change in the default parameters of any layer is done.

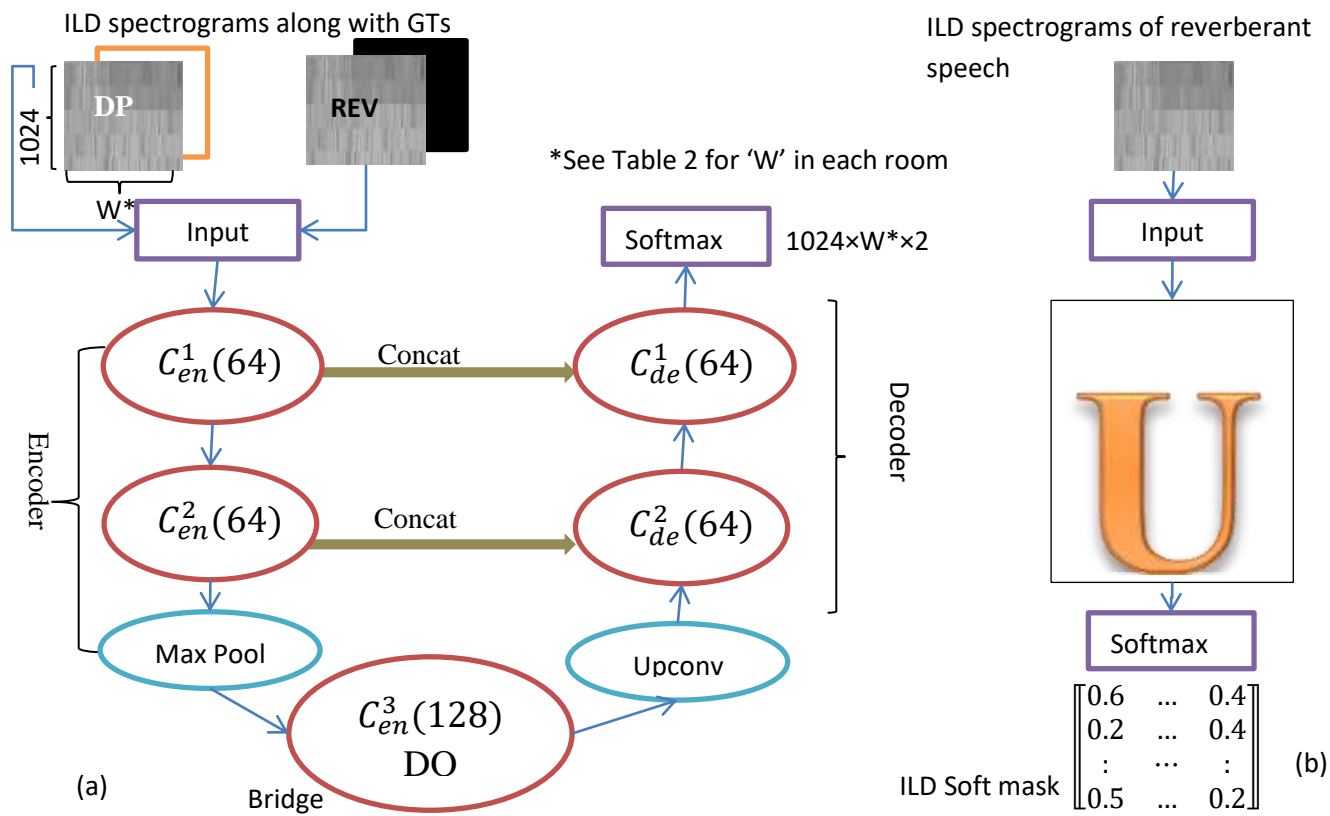

**Figure 5:** (a) Level U-Net architecture. Here only the convolution layers are shown. Number of feature maps in each convolution layer is shown inside the parenthesis in each circle. All convolution layers and the 'Upconv' layer are followed by ReLU (rectified linear unit) activation layer. The 'DO' stands for 'drop out', which is set to 50% by default. (b) U-Net after training i.e., during the prediction phase. The architecture is same as in (a)

The settings of hyper parameters, during training phase of the level U-Net, are mentioned in table 3.

**Table 3:** U-Net training parameters

| U-Net parameters | Values |
|---|---|
| Training samples per class | 12000 |
| Training and testing examples ratio | 0.98/0.02 |
| Input image dimensions (height × width) | STFT spectrogram size in table 2 according to $RT_{60}$ of room. |

| Encoder depth | 1 |
| --- | --- |
| Optimizer | sgdm (stochastic gradient decent method) |
| L2 regularization | 0.0001 |
| Momentum | 0.95 |
| Initial learning rate | 0.01 |
| Mini batch size | 8 |
| Number of epochs for $RT_{60}$ = 250ms | 30 |
| Number of epochs for $RT_{60}$ = 470ms | 13 |
| Number of epochs for $RT_{60}$ = 700ms | 10 |
| Number of epochs for $RT_{60}$ = 890ms | 5 |

The stopping criterion of training is taken as the leveling of the training accuracy curve.

### e) Ground truth (GT)

The ground truth (GT) of each class during training must have an image dimension equal to the STFT spectrogram size mentioned in table 2. So, the size of the ground truth varies with the room's $RT_{60}$ value. It consists of all the white pixels (pixel value = 255) for the DP class and all the black pixels (pixel value = 0) for the REV class.

### f) Comparative dereverberation algorithms

We have compared our proposed algorithm in four different rooms mentioned in table 2 with two dereverberation algorithms; the first from the 'classical' techniques, utilizing signal processing, the second utilizing the deep learning. A brief overview of these algorithms is given below.

#### 1. Weighted prediction error (WPE) [28]

The reverberation causes the lengthening of RIR. The algorithm proposed in [28], shortens this impulse response by using sub-band domain multi-channel linear prediction filters. Using these filters, the RIR shortening process is generalized and can work in all kinds of acoustic conditions. This method also works well for dereverberating acoustic mixtures collected at multiple microphones. The WPE algorithm is probably the most widely used algorithm for speech dereverberation. Many ASR studies report that WPE is the best classical signal processing models available, that suppresses reverberation with low speech distortions, and consistently improves the ASR performance, even for multi-conditionally trained ASR back-ends [29]

#### 2. Bidirectional long short term memory (BLSTM) based dereverberation model [29]

The dereverberation system proposed in [29] is a three stage dereverberation model using the bidirectional long short term memory (BLSTM) deep neural network, complemented by beamforming. The first stage consists of a BLSTM deep neural network; trained on real and imaginary (RI) components of the direct path waves. There are as many BLSTM trained networks as there are microphones installed, to collect the target signal. The purpose of this stage is to enhance the direct path signal received at each microphone. After enhancing through many such parallel BLSTM networks, the enhanced copies of microphone signals enter the second stage, which consists of a minimum variance distortionless response (MVDR) beamformer, which beamformed all single channel speeches, received from the first stage. The output of the beamformer is subtracted from the input signal (reverberant speech) at the first microphone in the time frequency (TF) domain to extract the RI components of the reverberations. These RI components of reverberations are then used for the training of another BLSTM network in the third stage of the model. This third stage trained network is then used to identify the reverberant RI components in the received signal. The block diagram of dereverberation network is shown in Figure 6.

The exact comparison of BENET with BLSTM dereverb model is not possible due to three main differences in training and testing conditions. The first difference is the type of noise used for the two models. Secondly, for model of [29], all training was done on data of [30], which was not available to us, nor were their RIRs, on which the training was done and thirdly, their target to microphone spacing was not similar to our arrangement, apart from the architectural differences of the two models. Also this model is not a binaural dereverberation model.

But, we chose to compare BENET with BLSTM model, as the design of our proposed model is very much inspired from this model. Like our proposed model, BLSTM based dereverberation system in [29] is also using the beamformer at the front, supported by the DNN at the back-end. However the number of beamformers in their network (1 vs.2), type of DNN (BLSTM vs. U-Net), number of DNN stages (2 vs. 1) and beamformer geometry (circular vs. linear) of [29] are different from BENET.

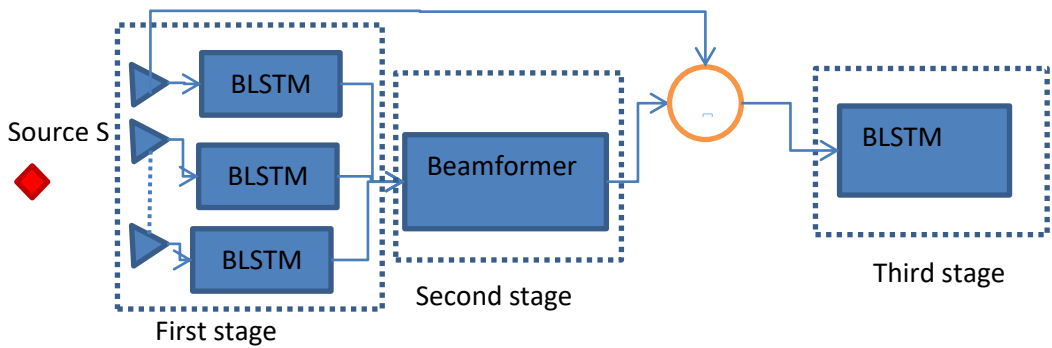

**Figure 6:** Dereverberation model of [29]

Due to limited computational resources available to us, we have mentioned the results of BLSTM algorithm directly from its paper, along with its respective acoustic conditions, under which the experiments were carried out.

### g) Evaluation criteria

The output of BENET is evaluated under noisy conditions using three evaluation metrics for comparison of different models. The evaluation metrics used are signal-to-reverberation modulation energy ratio (SRMR) [31], cepstral distance (CEP) [32] and frequency weighted segmental SNR (fwsegSNR) [33], as these metrics were also used for the evaluation of BLSTM model. Except for CEP, for all other metrics, higher is better. Except for SRMR, which does not require any reference, all other metrics use the direct path signal to the left microphone of binaural setup of Figure 2(c), as a reference for the performance evaluation.

### h) STFT parameters for creating interaural spectrograms

The STFT parameters used for converting the time domain signal to time frequency domain for the beamformer (equation (3)) and interaural spectrograms (equation (5), (6) and (7)) are summarized in table 4.

**Table 4:** STFT parameters for equations (3), (5), (6) and (7)

| Parameters | Eq. (3) | Eqs. (5), (6) and (7) |
|---|---|---|
| Sampling frequency | 16 kHz | 16 kHz |
| Window Shape | Hamming | Hamming |
| STFT frame length | 400 samples | 1024 samples |
| Hop size | 160 samples | 256 samples |
| FFT length | 512 | 1024 |

The non-target signal extracted from the reverberated signal at each beamformer is again converted to time domain audio signal by overlapadd function in voicebox [34].

### i) Beamformer comparison

We have tested different beamformers for BENET under similar acoustic conditions. Finally we have chosen minimum variance distortionless response (MVDR) [35] for two reasons. The most important one is its better performance in experiments on a small amount of traininig dataset and the second is that, it is also used in BLSTM model. Eight (8) elements were kept in each beamformer and no other changes were done in the model of [35]. During the training phase, the source is placed at the center of the two beamformers, as shown in Figures 4(a) and (b), making an angle of -10.758$^o$ with the perpendicular plane passing through the center of left beamformer and +10.758 $^o$ with the perpendicular plane passing through the center of right beamformer. The geometry and spacing of sensors, the room dimensions and the source position are given as input to simulator of [25], to achieve RIRs according to these steering angles. The steering angle of each beamformer is focused on the real source $S$. The REV signal from each beamformer is obtained by subtracting the beamformed signal from the reverberated speech collected at the first sensor of each beamformer, which has also collected an omnidirectional signal from multiple virtual sources, as the single sensor itself is omnidirectional. The process of subtraction is depicted mathematically in equation (3) of Section 3.

### j) Noisy conditions

In noisy conditions, we have performed the experiments under the signal to noise ratio (SNR) of 20 dB to make a fair comparison with BLSTM model, as this SNR is also used for their model testing. However, as the model of [29] has added their own recorded fan sound, which was not available to us, so, we have added white noise to our speech signal of [26] by [36]. One thing must be clear, that this noisy data is not used during the training phase. It is only used for the prediction phase.

## 5. Experiments and Results

We have performed two different experiments. The details and results of these experiments are given below.

Case I: **BENET comparison with other algorithms.**

In this experiment, we have compared the output of BENET model of Figure 2(c) with the other dereverberation algorithms mentioned above.

For all comparisons, the tests are carried under all reverberant conditions mentioned in table 2 at SNR of 20dB, with 5 reverberant audio speech signals in each room, generated from the clean files of two seconds each, after convolving them by the RIRs of these rooms and the results of all the rooms are then averaged.

First, the source is placed at the same position, as in the training phase. But the results were not encouraging, so we have tested the model at positions, in the vicinity of its training position and the BENET's performance is much improved as compared to its previous performance at the training position.

The average results of testing our proposed model in noisy conditions, inside a room with $RT_{60}$ of 890ms, when placed at and other slightly shifted positions in elevation (up and down), azimuth (left and right) and front and back in the range of [5cm:5cm:15cm] of the training position, are shown in Table 5 and depicted in Figure 7 .

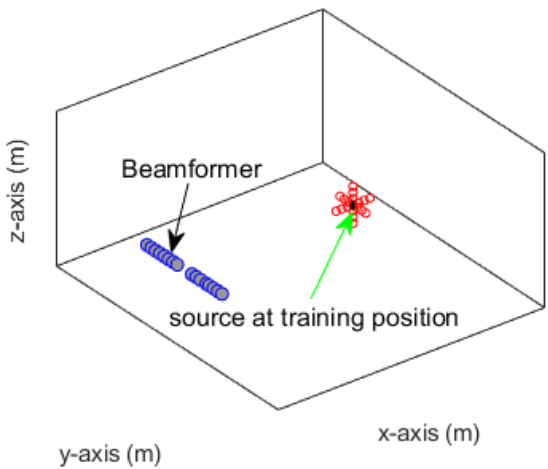

**Figure 7:** Source positions in the vicinity of training position during the testing phase are shown by the red dots. The training position itself is shown by the black dot.

**Table 5:** Performance of BENET trained in $RT_{60}$ (s) =890ms at different positions during the testing phase

| Position of source during testing | CEP | fwsegSNR (dB) | SRMR (dB) |
|---|---|---|---|
| Same as the training position | 3.02 | 10.3 | 3 |
| Up/down shift | 1.0 | 20.1 | 3.1 |
| Left/right shift | 1.1 | 19.8 | 2.9 |
| Front/back shift | 1.1 | 19.7 | 3.2 |

Although the SRMR after shifting the source remains almost similar to the one obtained at its training position, yet there is an enormous improvement in CEP and fwsegSNR. The exact cause of this improvement needs further investigation but it appears that moving the source slightly in the vicinity of its training position has not altered its spatial cues much [21], but has changed the reverberation pattern inside room, resulting in creation of new virtual sources and diminishing of the older ones, resulting in misclassification of reverberation spatial cues, adding the early reverberation more in the direct path signal, thus enhancing its intelligibility, as depicted by an increase in the fwsegSNR, as this metric is found to correlate well with speech intelligibility [37]. So, it is suggested that the source must be placed in the vicinity and not at the training position during the testing phase, for better system performance.

Our proposed system has worked well on the positions in the vicinity of its training position, allowing mild excursions of source during his speech. Moving far apart the training positions results in changing the spatial cues of the real source and consequently their misclassification and the resultant drop in performance as shown in Table 6, where the source placed at 5cm to the left of its training position is compared with the source placed at 50 cm towards left of the training position, inside a room with $RT_{60}$ =890ms.

**Table 6:** Performance of BENET at near and far positions with respect to its training position

| Position of source during testing | CEP | fwsegSNR (dB) | SRMR (dB) |
|---|---|---|---|
| 5cm left of training position | 1.1 | 20 | 3.1 |
| 100cm left of training position | 1.4 | 18 | 2.6 |

The results in Table 6 show that moving farther away from the trained position results in degradation of systems performance. This is because the system is spatial cue-based and moving the source farther would result in changing its spatial cues too much, resulting in them being unrecognized by the trained system.

Now we will compare our proposed system with other dereverberation algorithms. The average results of source placed 10cm in front of its training position in BENET, with the other dereverberation algorithms, are mentioned in Table 7.

**Table 7:** Comparison of BENET with other algorithms under noisy conditions

| Method | $RT_{60}$ (s) | CEP | fwsegSNR (dB) | SRMR (dB) |
|---|---|---|---|---|
| **BENET** | 0.25, 0.47, 0.7 | **1.74** | **17** | 3.5 |
| **WPE** | 0.25, 0.47, 0.7 | 3.13 | 9 | 4.1 |
| **BLSTM** | 0.25, 0.5, 0.7 | 3.01 | 16.94 | **6.38** |

Under similar $RT_{60}$ values, Table 7 shows that BENET has outperformed other competing models in terms of CEP and fwsegSNR. The lower value of CEP is useful for machine listening applications, which require minimum distortion, as machines are not as resilient as the human listening. However, the SRMR of our proposed algorithm is slightly lower than WPE (<1 dB), and it is almost half in value of that offered by BLSTM [29]. The two reasons behind this poor performance may be, i) the differences in the room design and RIRs of [29] from those, that we have used for our proposed model and ii) the effect of keeping the beamformers intact in [29], also during its testing phase (as discussed in context of Table 8).

Although, speech-to-reverberation modulation energy ratio (SRMR-CI) [38] (the objective metric designed especially to measure the speech intelligibility for CI users in noisy and reverberant environments), is not mentioned in paper [29], we found that the average SRMR-CI of our proposed algorithm is 0.1 dB higher than WPE (4.3dB vs. 4.2dB), under the acoustic conditions mentioned in Table 7, verifying its suitability for people, using the restorative hearing instruments.

The average performance of our proposed model is better than WPE, as it is equipped with both the beamforming and the DNN. Although the same is true for the BLSTM model, yet BENET has surpassed it, not only in performance, but also in terms of requiring lesser training data (80 hours for BLSTM vs. 12 hours for BENET). However, our data set creation step is very lengthy due to involvement of beamformers, while they have trained their system on the already available dataset of REVERB challenge [30]. The reason of their better SRMR is because of using the beamforming even in the testing phase, while our proposed algorithm uses only two microphones during the testing phase. It has been found that beamforming improves the SRMR of BENET also, when it is used in its testing phase too, as shown in Table 8.

**Table 8:** BENET performance with and without using beamformers in the testing phase

| Method | Microphones in the testing phase | $RT_{60}$ (s) | CEP | fwsegSNR (dB) | SRMR (dB) |
|---|---|---|---|---|---|
| **BENET** | Two microphones | 0.25, 0.47, 0.7 | **1.7** | **17** | 3.5 |
| **BENET** | Two beamformers | 0.25, 0.47, 0.7 | 1.7 | 14.1 | 4.4 |

As clear from Table 8, keeping the beamformers intact during the testing phase has degraded the fwsegSNR, yet it has improved the SRMR by more than 1 dB. But, these variations in SRMR and fwsegSNR must be compromised according to their impact on the application under consideration. In the next experiment, we will check the generality of BENET model in unseen acoustic conditions.

Case 2: ***Testing BENET in unseen acoustic conditions***

Ideally, all binaural dereverberation must be implemented without any prior measurement of the room response, thus being blind or at least semi-blind (when some broad parameters related to the acoustic environment have to be known) [1]. In this experiment, we will test BENET trained in one room, in rest of other rooms, where it was not trained (unseen conditions). For testing the network trained in low reverberation conditions, we have selected the BENET model trained in room of $RT_{60}$ =250ms, and test it under the medium (470ms) and high reverberation (890ms) conditions. Likewise, we have selected the model trained in high reverberation (890ms) and test it in other two unseen conditions of 250 and 470ms. The results are shown in Tables 9 and 10. The source is placed 10cm in front of its training position during this experiment.

**Table 9:** Performance of BENET trained in $RT_{60}$ (s) = 250ms in medium and high reverberant conditions.

| $RT_{60}$ (s) | CEP | fwsegSNR (dB) | SRMR (dB) |
|---|---|---|---|
| 0.25 (seen) | 2.6 | 11.4 | 5.3 |
| 0.47 (unseen) | 1.1 | 20 | 5.4 |
| 0.89 (unseen) | 2.5 | 10.7 | 3.4 |

**Table 10:** Performance of BENET trained in $RT_{60}$ (s) = 890ms in medium and high reverberant conditions.

| $RT_{60}$ (s) | CEP | fwsegSNR (dB) | SRMR (dB) |
|---|---|---|---|
| 0.89 (seen) | 1.1 | 19.8 | 3.4 |
| 0.47 (unseen) | 1.1 | 20.7 | 5.4 |
| 0.25 (unseen) | 1.8 | 15.3 | 7.1 |

As shown in Table 9, BENET trained in low reverberations ($RT_{60}$ (s) = 250ms) maintains its performance in unseen conditions of medium reverberations ($RT_{60}$ (s) = 470ms), but there is a sharp decline in all metrics in high reverberant condition (890ms). Likewise, as shown in Table 10, the BENET trained in higher reverberant conditions ($RT_{60}$ (s) = 890ms) failed to adapt to the lower reverberant condition ($RT_{60}$ (s) = 250ms). Except SRMR, all performance metrics decline. So, it can be concluded that our proposed models works well under unseen acoustic conditions, which are closer to the one under which the model was trained, while the performance drops in unseen conditions which are entirely different from the training conditions.

## 6. Conclusion

In this paper, our main idea is to preserve the beamforming effect in a deep neural network (U-Net), so that after training, it can discriminate and identify the direct path spatial cues from the reverberations by using only the binaural setup i.e. without the aid of beamformers. Our proposed model, surpasses the signal processing and deep learning dereverberation models, when during the testing phase, the source is placed in the vicinity of its original position of training phase. It's lower CEP than the competing algorithms (considered in this paper), makes it an ideal choice for ASR and ASV applications. The higher values of fwsegSNR and SRMR-CI of our proposed model, shows improved intelligibility, both for normal listeners and for users of assistive listening devices (e.g. cochlear implants). Also the performance of the proposed system is not depreciated under relatively similar unseen acoustic conditions. Although the proposed model has shown its effectiveness with simulated RIRs, yet, in future it requires testing with the real RIRs, pretrained networks and implementation using the light weight U-Net architecture (e.g. as used by speech separation model [39]), suitable for wearable and mobile devices, before being incorporated in real life dereverberation applications.

**Funding**

This work is funded by Higher Education Commission (HEC), Pakistan, under project no. 6330/KPK/NRPU/R&D/HEC/2016.

**Acknowledgements**

We gratefully acknowledge the support of NVIDIA Corporation with the donation of the Titan X Pascal used for this research. We would also like to thank the AI in health care lab, University of Engineering and Technology, Peshawar for sharing their computational facilities.

**References**

[1]. Blauert, Jens.: "The technology of binaural listening", Springer publications, 2013.

[2]. Patrick A. Naylor, Nikolay D. Gaubitch.: "Speech dereverberation", Springer publications, 2010.

[3]. Ivan Tashev, Henrique Malvar.: "System and method for beamforming using a microphone array", Patent No. US7415117B2, United States.

[4]. Mandel Michael I, "Binaural Model-Based Source Separation and Localization", Ph.D. thesis, Columbia University, 2010.

[5]. Habets: "Speech dereverberation using statistical reverberation models", Signals and Commmunication Technology, Springer, London, 2010.

[6]. Jitendra D. Rayala, Krishna Vemireddy.: "Real-time microphone array with robust beamformer and postfilter for speech enhancement and method of operation thereof". Patent No. US9538285B2, United States.

[7]. Syed Mohsin Naqvi, Muhammad Salman Khan, Qingju, Jonathon Chambers.: "Multimodal blind source separation with a circular microphone array and robust beamforming", Proc. EUSIPCO, Spain, 2011.

[8]. May, Tobias.: "Robust speech dereverberation with a neural network-based post-filter that exploits multi-conditional training of binaural cues." IEEE/ACM Transactions on Audio, Speech, and Language Processing, Vol . 26(2), pp 406-4142017.

[9]. Westermann, Adam, Jörg M. Buchholz, and Torsten Dau.: "Binaural dereverberation based on interaural coherence histograms." Journal of the Acoustical Society of America, Vol. 133(5) pp 2767-2777, 2013.

[10]. Allen, J. B., Berkley, D. A., and Blauert, J. .: "Multimicrophone signal-processing technique to remove room reverberation from speech signals," Journal of the Acoustical Society of America, Vol. 62, 1977, pp 912–915.

[11]. Beit-On, Hanan, and Boaz Rafaely.: "Binaural direction-of-arrival estimation in reverberant environments using the direct-path dominance test", Proc. International congress on acoustics, Germany, Sept. 2019.

[12]. K. Lebart, J. Boucher, and P. Denbigh.: "A new method based on spectral subtraction for speech

Dereverberation",. Acta Acust./Acustica, Vol . 87, pp 359–366, 2001.

[13]. H. W. Löllmann and P. Vary.: "Low delay noise reduction and dereverberation for hearing aids", EURASIP Journal Advances in Signal Processing, pp 1–9, 2009.

[14]. M. Jeub, M. Schafer, T. Esch, and P. Vary.: "Model-based dereverberation preserving binaural Cues", IEEE/ACM Transactions on Audio, Speech, and Language Processing, Vol. 18(7), pp 1732–1745, 2010.

[15]. J. H. Lee, S. H. Oh, and S. Y. Lee.: "Binaural semi-blind dereverberation of noisy convoluted speech signals", Neurocomputing, Vol. 72, pp 636–642, 2008.

[16]. Li, Ruwei, et al.: "Speech separation based on reliable binaural cues with two-stage neural network in noisy-reverberant environments." Applied Acoustics 168 (2020): 107445.

[17]. Arifianto, Dhany, and Mifta N. Farid.: "Dereverberation binaural source separation using deep learning", Journal of the Acoustical Society of America, Vol. 144.3, pp. 1684-1684, 2018.

[18]. K. Tan, B. Xu, A. Kumar, E. Nachmani and Y. Adi.: "SAGRNN: Self-Attentive Gated RNN For Binaural Speaker Separation With Interaural Cue Preservation," in *IEEE Signal Processing Letters*, Vol. 28, pp. 26-30, 2021.

[19]. Kazushi Nakazawa, Kazuhiro Kondo.: "De-Reverberation using CNN for Non-Reference Reverberant Speech Intelligibility Estimation", Proc. International Congress on Acoustics, Aachen, Germany, September 2019.

[20]. Hanwook Chung, Vikrant Singh Tomar and Benoit Champagne.: "Deep convolutional neural network-based inverse filtering approach for speech de-reverberation", arXiv:2010.07895v1 [cs.SD] 15 Oct 2020.

[21]. Sania Gul, Muhammad Sheryar Fulaly, Muhammad Salman Khan, Syed Waqar Shah.: "Clustering of spatial cues by semantic segmentation for anechoic source separation", Applied Acoustics, Vol. 171, January 2021.

[22]. Sania Gul, M. S. Khan, and S. W. Shah.: "Integration of deep learning with expectation maximization for spatial cue based speech separation in reverberant conditions", Applied Acoustics, Vol. 179, August 2021.

[23]. Michael I. Mandel, Ron J. Weiss, Daniel P. W. Ellis.: "Model-based expectation-maximization source separation and localization", IEEE transactions on audio, speech and language processing, Vol. 18, No. 2, February 2010.

[24]. Parham Aarabi.: "Self-localizing dynamic microphone arrays", IEEE Transactions on Systems, Man, And Cybernetics—Part C: Applications and Reviews, Vol. 32, No. 4, 2002.

[25]. Eric A. Lehmann (2020). Image-source method for room impulse response simulation (room acoustics) (https://www.mathworks.com/matlabcentral/fileexchange/20962-image-source-method-for-room-impulse-response-simulation-room-acoustics), MATLAB Central File Exchange.

[26]. "DAPRA TIMIT acoustic phonetic continuous speech corpus", http://www.ldc.upenn.edu/Catalog/LDC93S1.html

[27]. http://www-mmsp.ece.mcgill.ca/Documents/Data/

[28]. Takuya Yoshioka and Tomohiro Nakatani.: "Generalization of multi-channel linear prediction methods for blind MIMO impulse response shortening", IEEE Transactions on Audio, Speech and Language Processing, Vol. 20, No. 10, December 2012.

[29] Zhong-Qiu Wang and Deliang Wang.: "Deep Learning based target cancellation for speech deverberation", IEEE/ACM transactions on audio, speech and language processing, Vol. 28, 2020.

[30]. Keisuke Kinoshita et al.: "A summary of the REVERB challenge: State-of-the art and remaining challenges in reverberant speech processing research," EURASIP J. Adv. Signal Process., Vol. 7, No. 1, 2016.

[31]. https://github.com/MuSAELab/SRMRToolbox

[32]. https://github.com/schmiph2/pysepm

[33]. Zexin Liu, Heather T. Ma and Fei Chen.: "A new data-driven band-weighting function for predicting the intelligibility of noise-suppressed speech", Proc. APSIPA Annual Summit and Conference, December 2017, Malaysia.

[34]. http://www.ee.ic.ac.uk/hp/staff/dmb/voicebox/voicebox.html

[35]. Takuya Higuchi, Nobutaka Ito, Takuya Yoshioka and Tomohiro Nakatani.: "Robust MVDR beamforming using time-frequency masks for online/offline ASR in noise", Proc. ICASSP, 2016. https://github.com/snsun/cgmm_mvdr

[36]. https://www.mathworks.com/matlabcentral/fileexchange/33198-segmental-snr?focused=5202375&tab=function

[37]. Z. Liu, H. T. Ma and F. Chen.: ''A new data-driven band-weighting function for predicting the intelligibility of noise-suppressed speech", Proc. APSIPA Annual Summit and Conference, Malaysia, December 2017.

[38]. Santos, João F., et al.: "Objective speech intelligibility measurement for cochlear implant users in complex listening environments." Speech communication 55.7-8 (2013): 815-824.

[39]. K. M. Jeon, C. Chun, G. Kim, C. Leem, B. Kim and W. Choi.:" Lightweight U-Net Based Monaural Speech Source Separation for Edge Computing Device", Proc. *IEEE International Conference on Consumer Electronics (ICCE)*, Seoul, Korea, Jan. 2020.